# The binary primes sequence for computational hardening of pseudorandom sequences

B. Prashanth Reddy & Subhash Kak

***Abstract*** - This paper proposes the use of the binary primes sequence to strengthen pseudorandom (PN) decimal sequences for cryptography applications. The binary primes sequence is added to the PN decimal sequence (where one can choose from many arbitrary shift values) and it is shown that the sum sequence has improved autocorrelation properties besides being computationally hard. Also, an analysis on the computational complexity is performed and it is shown that the complexity for the eavesdropper is of the order $O(N^N)$ and therefore, the proposed method is an attractive procedure for cryptographic applications.

**Introduction**

In decentralized network architectures and electronic currency protocols, one needs random sequences with good complexity and correlation properties. This paper examines the properties of the binary primes sequence and proposes their use for strengthening of pseudorandom sequences for generating keys and for signing secure protocols. The paper presents a method to exploit the complexity of the primes sequence and convert that to the randomness of the binary sequence.

The reason why this may not have been examined before is that the primes come along rather slowly with the integer value as we know from the prime number theorem [1]. Our idea to deal with this problem is to use several shifted versions of the binary primes sequence so that the number of 0s and 1s are approximately balanced.

Pseudonoise (PN) sequences are random binary sequences with a maximum period which are generated by using a linear feedback shift registers [2]-[4]. PN sequences are widely used for generating keys even though they are not strong cryptographically. The class of pseudorandom decimal (D) sequences [5] are sometimes used in place of shift register sequences:

$$a(i) = (2^i \bmod q) \bmod 2 \qquad\qquad (1)$$

where q is a prime number. Generated D sequences possess good autocorrelation properties although they are also cryptographically weak.

We show in this paper that adding a suitable binary primes sequence to the D sequence leads to excellent correlation characteristics of the resultant sequence, which can, therefore, be used in a variety of applications that require random sequences. The binary primes sequence may likewise be added to the PN sequence for computational hardening [6].

The next section presents the preliminaries regarding the binary primes sequence. Subsequent sections investigate this and the composite pseudorandom sequences for their randomness properties.

**The binary primes sequence**

The binary primes sequence obtained by the mapping:

$$b(k) = \begin{cases} 1, & \text{if } k = p, \quad \text{a prime number} \\ 0, & \text{if } k \neq p, \quad \text{not a prime number} \end{cases} \qquad (2)$$

It is understood that b(k) =0, for $k \neq 0$.

Since the problem of computing the next primes sequence is computationally intensive, the primes sequence is intuitively random. However, it is not balanced in its distribution of 0s and 1s. Now according to the Prime Number Theorem [6], the number of prime numbers less than an integer *N* is given by

$$\Pi(N) \sim \frac{N}{\ln N} \qquad (3)$$

When we consider first 1000 natural numbers, the number of primes less than 1000 is about 144 or the binary sequence has approximately 144 1s and 856 0s. We propose to balance this sequence by adding several shifted versions of the primes sequence.

The number of shifters required to generate the binary primes sequence will be

$$\frac{N/2}{\frac{N}{\ln N}} = \frac{1}{2} \ln N \qquad (4)$$

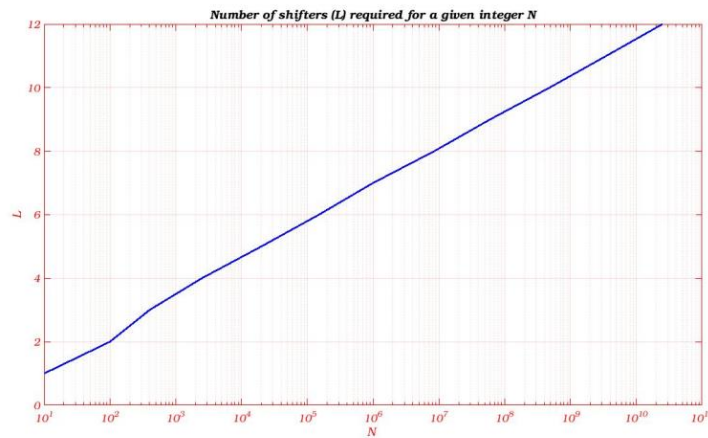

Figure 1: Number of shifters L required for a given integer N

Table 1: Prime sequences with one shift for *n*=10 **(L=1)**

| n = 10 | 1 | 2 | 3 | 4 | 5 | 6 | 7 | 8 | 9 | 10 | 1s count |
|--------|---|---|---|---|---|---|---|---|---|----|----------|
|        | 0 | 1 | 1 | 0 | 1 | 0 | 1 | 0 | 0 | 0  | **4 (a₀=0)** |
|        | 0 | 0 | 1 | 1 | 0 | 1 | 0 | 1 | 0 | 0  | *4 ➜ shift #1* **(a₁=1)** |
|        | 0 | 1 | 0 | 1 | 1 | 1 | 1 | 1 | 0 | 0  | **6** |

Table 2: Prime sequences with two shifts for *n*=10 **(L=2)**

| n = 10 | 1 | 2 | 3 | 4 | 5 | 6 | 7 | 8 | 9 | 10 | 1s count |
|--------|---|---|---|---|---|---|---|---|---|----|----------|
|        | 0 | 1 | 1 | 0 | 1 | 0 | 1 | 0 | 0 | 0  | **4 (a₀=0)** |
|        | 0 | 0 | 1 | 1 | 0 | 1 | 0 | 1 | 0 | 0  | *4 ➜ shift #1* **(a₁=1)** |
|        | 0 | 0 | 0 | 1 | 1 | 0 | 1 | 0 | 1 | 0  | *4 ➜ shift #2* **(a₂=2)** |
|        | 0 | 1 | 0 | 0 | 0 | 1 | 0 | 1 | 0 | 0  | **3** |

Tables 1 & 2 show the use of one and two shifters to produce a balanced primes sequence for n=10. Note that shifters employed here are shifting the prime sequence, b(k) right by one unit. However, a shifter can have variable shifts in practical.

The generalized binary primes sequence, $B_{N,L}(k)$ is sum of several binary primes sequences of shifts $a_0$, $a_1$ …. $a_{L-1}$ units, where $a_0=0$, is the unshifted version of b(k).

$$B_{N,L}(k) = \sum_{i=0}^{L-1} b(k - a_i) \tag{5}$$

**Autocorrelation of $B_{N,L}(k)$**

The autocorrelation function C(n) for the binary primes sequence $B_{N,L}(k)$ is given by:

$$C(n) = \frac{1}{N} \sum_{m=0}^{N-1} B_{N,L}(m) . B_{N,L}(m + n) \tag{6}$$

where L is the number of shifters and N is the period of b(k).

Figure 2 below shows the autocorrelation function of the binary primes sequence $B_{N,L}(k)$ by considering a prime number 997. Here terms with $a_1 = 11$, $a_2 = 77$, $a_3 = 111$ of b(k) are added to b(k) to generate B(k) sequence. The absolute non-zero value of c(n) is noticed to be 0.3133.

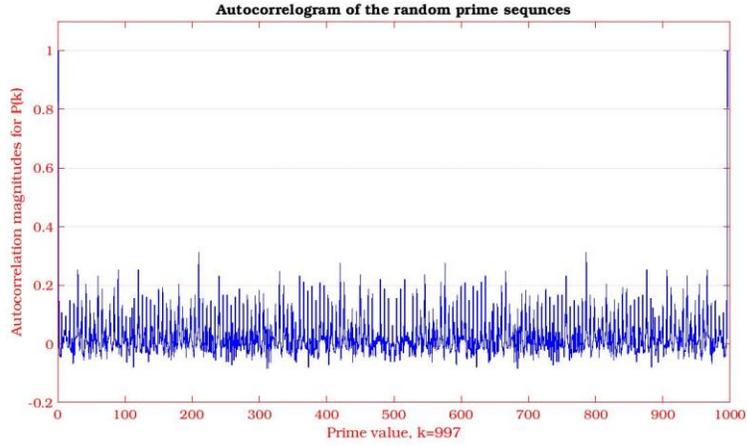

Figure 2: Autocorrelation graph of the Binary primes sequence $B_{N,L}(k)$ for a prime number 997

**Randomness measure for $B_{N,L}(k)$:**

The randomness measure for the binary primes sequence, $B_{N,L}(k)$, is given by the metric below [7]:

$$R_N(B) = 1 - \frac{\sum_{n=1}^{N-1}|c(n)|}{N-1} \qquad (7)$$

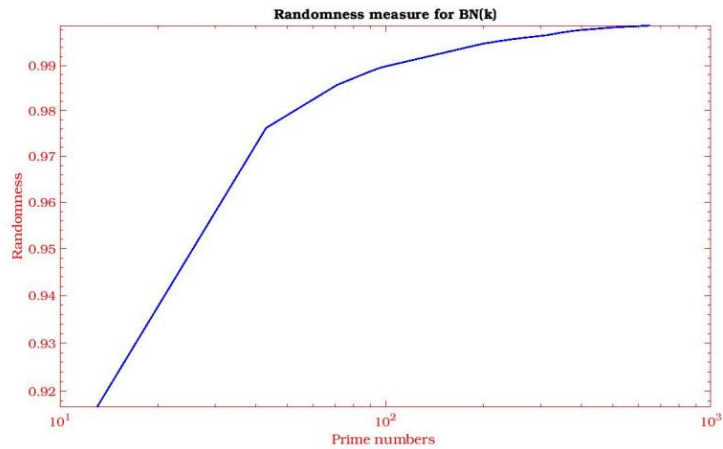

Figure 3: Binary Prime sequences performances in terms of randomness

The randomness measure of the sequences of different length with the same shift factors that is shown in Figure 3 confirms that the randomness metric increases as the period increases. The randomness measure of $B_{N,L}(k) = B_{199,3}(0,7,11,22) = 0.9949$.

**Computational hardening of PN sequences**

We consider adding the D sequence, $a(k)$ from equation (1), to the binary primes sequence $B_{N,L}(k)$. Let the sum sequence be represented by $P(k)$:

$$P(k) = B_{N,L}(k) + a(k) \qquad (8)$$

**Autocorrelation function for the advanced version of binary primes sequence P(k):**

The autocorrelation function of P(k) is calculated by C(k) = $\frac{1}{N}\sum_{m=0}^{N} P(m).P(m+k)$ where N is the period for the maximum length periodic sequence.

Figure 4 shows the autocorrelation graph of P(k) for the prime number 199. $B_{N,L}(K) = B_{199,3}(0,7,11,22)$=b(k) + b(k-7) + b(k-11) + b(k-22) and in Figure 5 shows the autocorrelation graph of P(k) for the prime number 997. $B_{N,L}(K) = B_{997,3}(0,11,77,111)$=b(k) + b(k-11) + b(k-77) + b(k-111) and in both cases the results are excellent.

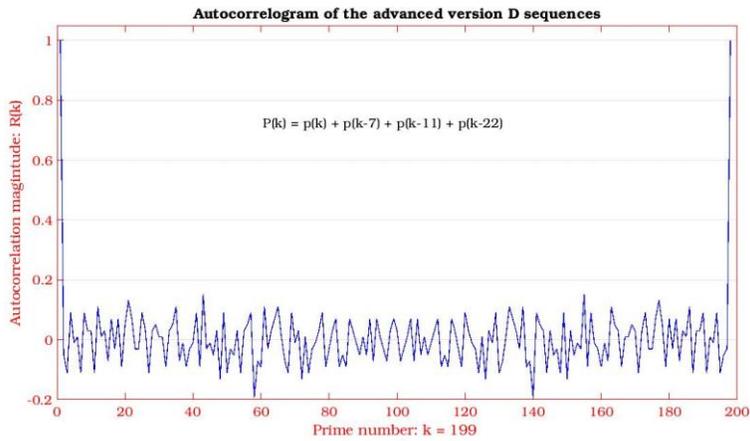

Figure 4: Autocorrelation of P(k) for prime number 199

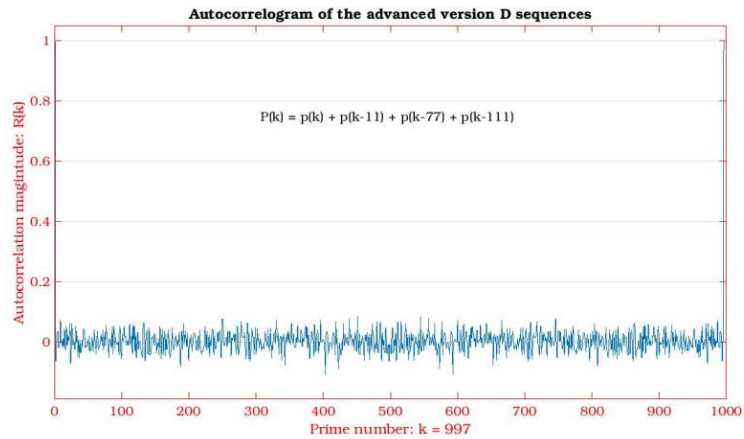

Figure 5: Autocorrelation of P(k) for prime number 997

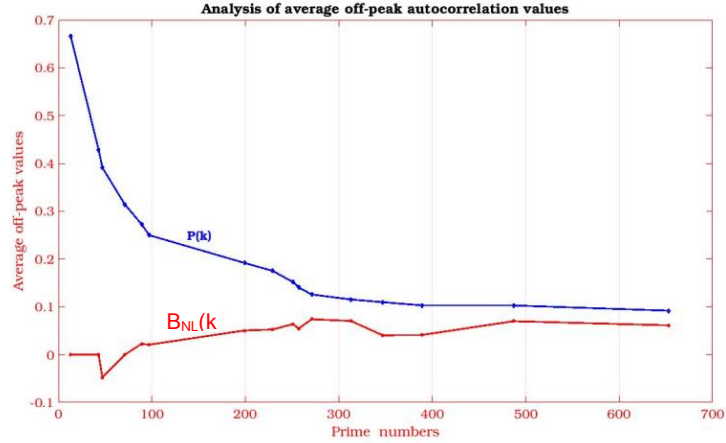

Figure 6: comparison of average off-peak curves for $B_{N,L}(k)$ and P(k)

Figure 6 presents the average off-peak magnitudes of P(k) and binary primes sequence $B_{N,L}(k)$ for different prime numbers ranging between 40 to 650. We see that the quality of randomness increases with p.

**Complexity analysis**

From the point of view of the eavesdropper the determination of the $B_{N,L}(k)$ sequence offers exponental complexity. The eavesdropper must try out not only different values of N and L in the binary primes sequence but also the variables $a_0, a_1, \ldots a_{L-1}$. Since the $a_i$s can be arbitrary values, the task of the eavesdropper is well high impossible.

According to the attacker's point of view, there are 3 unknowns factors in the given algorithm.

1. The prime number used to generate the random sequence
2. The number of shifters that have been utilized to generate the binary primes sequence, $B_{N,L}(k)$
3. Different shifts factors used to generate the primes sequence b(k)

If the binary primes sequence is generated by employing, a large value integer, N, then the complexity for an attacker is $\frac{N^2}{2 \ln N} \cdot N^{\frac{N}{\ln N}}$

For example if we consider, N ~ $10^6$. Then the complexity order will be $\frac{10^{12}}{2 \ln 10^6} \times 10^{\frac{6 \times 10^6}{\ln 10^6}}$.

Recounting the situation for the general case, we have the following:

- The number of prime numbers less than an integer n is given by $\frac{N}{\ln N}$,
- the number of shifters required is $\sim \frac{N/2}{\frac{N}{\ln N}} = \frac{1}{2} lnN$

- different shift factors provided to the primes sequence, (dependent on the number of shifters employed) is given by $N^{\frac{N}{\ln N}}$

Thus total number of possibilities as far as the attacker should try in order to guess the binary primes sequence is given by ➔ $\frac{N^2}{2 \ln N} \cdot N^{\frac{N}{\ln N}}$

The effort in guessing all the above compuational factors cannot be easy to the attacker due to the complexity of the algorithm.

**Conclusion**

Although this paper used binary primes sequence to strengthen decimal sequences, the method can be likewise employed to strengthen other pseudorandom sequences. We showed that the autocorrelation properties of the resultant sequences are excellent and we related the variation of average off-peak value of it to change in prime numbers. The cost of adding the binary primes sequence is minimal because it can be precomputed and stored in the application and the user merely chooses the shift values that are to be employed. One can also use the shift values to represent a key that may be shared between two users by employing a suitable key-exchange protocol.

In conclusion, the effectiveness of the method of adding the binary primes sequence to a given PN sequence can be used for key generation and in other cryptographic applications. This algorithm will increase the computational complexity for the attacker and so this algorithm can potentially be applied in various cryptographic applications. Moreover, the binary primes sequences can also be employed to strengthen other pseudorandom sequences like the ones generated by shift registers.